\documentstyle[aps,preprint,12pt]{revtex}
\input{epsf.tex}

\newcommand{\be}{\begin{equation}}
\newcommand{\ee}{\end{equation}}
\newcommand{\bear}{\begin{eqnarray}}
\newcommand{\eear}{\end{eqnarray}}

\begin{document}
\textheight 7.5in
\topmargin 0.5in

\baselineskip 0.8cm

%\twocolumn[\hsize\textwidth\columnwidth\hsize\csname @twocolumnfalse\endcsname
%{\today}
\vspace*{0.5cm}

\noindent
{\bf Classification:} Biological Sciences, Biophysics\\

\vspace*{-0.5cm}

%\title{

\vspace*{0.5cm}
\noindent
{\bf Title:} {Protein threading  by learning } 

\vspace*{0.5cm}
%\author{

\noindent
{\bf Authors:}\\
{\sc Iksoo Chang${}^{1,2}$, 
Marek Cieplak${}^{1,3}$, 
Ruxandra I. Dima${}^4$, 
Amos Maritan${}^5$,
and
Jayanth R. Banavar${}^1$ 
}                   

\vspace*{0.5cm}
%\address
\noindent
{${}^1$Department of Physics, 
104 Davey Laboratory, The Pennsylvania State University, 
University Park, PA 16802, USA} \\
%\address
{${}^2$Department of Physics, Pusan National University, 
Pusan 609-735, Korea}\\
%\address
{${}^3$Institute of Physics, Polish Academy of Science, 
02-668 Warsaw, Poland} \\
%\address
{${}^4$Institute for Physical Science and Technology and Department of 
Chemistry and Biochemistry, University of Maryland, College Park, 
Maryland 20742, USA}\\
%\address
{${}^5$International School for Advanced Studies (SISSA) and
Abdus Salam International Center for Theoretical Physics, 
Via Beirut 2-4, 34014 Trieste, Italy,
and Instituto Nazionale di Fisica della Materia, Italy}\\

\noindent
{\bf Corresponding author:} Jayanth R. Banavar, 104 Davey Laboratory,
The Pennsylvania
State University, University Park, Pennsylvania 16802, phone: 814-863-1089,
FAX: 814-865-0978, email: jayanth@phys.psu.edu\\

{\small
\begin{description}
\item pages: 14 \vspace*{-0.5cm}
\item figures: 4  \vspace*{-0.5cm}
\item tables: 1 and 2 supplementary tables \vspace*{-0.5cm}
\item abstract: 74 words \vspace*{-0.5cm}
\item paper: 40,435 characters 
\end{description}      }

%\maketitle
%\begin{abstract}

\newpage
\vspace*{3cm}
\noindent
{\large {\bf Abstract}   }

\vspace*{0.5cm}
\noindent
{\bf
Using techniques borrowed from statistical physics and neural networks, we
determine the parameters, associated with a scoring function,
that are chosen optimally to ensure
complete success in threading tests in a training set of 
proteins. 
These parameters provide a
quantitative measure of the propensities of amino acids to be buried or exposed and
to be in a given secondary structure and are a good starting point for solving
both the threading and design problems.  
}

%\end{abstract}

%\pacs{PACS numbers:   } 

%% FOR TWO COLUMN ACTIVATE THE LINE BELOW
%]

\baselineskip 0.8cm
\parindent 0.75cm
\newpage

\narrowtext

The principal objective of this paper is a demonstration 
of the viability of a framework,
based on ideas from statistical physics and
neural networks, for attacking the protein threading problem.
Our work points to the difficulty associated with a commonly used
statistical procedure for determining such parameters.
We present the results of threading and design tests
and present a singular value decomposition (SVD) analysis of the parameters
which elucidate
the interplay between degree of burial
and secondary structure propensities in the folding problem.

The challenge of the protein folding problem (1-5)
is to deduce the native
state structure and thence the functionality of a protein from the 
knowledge of the sequence of amino acids. 
The successful completion of the human genome project has heightened
interest in this problem. The information readily available as input
are the sequences and native structures of a few
thousand proteins (6). Given an entirely new sequence, one needs to have
a sound strategy for determining its native state structure. 
A simpler problem, threading (7),
relies on the belief that the total number of
distinct folds in nature is only a few thousand (8) and attempts
to match the new sequence with the best among a selection of possible 
native state structures. (A difficulty associated with threading is that
due to steric constraints, one may not be able to mount a given
sequence on a piece of a native structure of a different sequence.
See, for example ref. 9) 
In order to assess the fit of a given sequence with a putative native
state structure, one might use a coarse grained representation of the amino
acids in a sequence and
postulate a scoring function with a simple functional form.
Perhaps the simplest such function 
is one which characterizes the propensities of 
the various types of amino acids to be in different environments:
\begin{equation}
S(\bar{s},\bar{\Gamma}) = \sum_{i} \sum_{m} n(i,m) \; \epsilon(i,m) \;\;,
\end{equation}
where $S$ is the score function which is a measure of the match of 
a sequence, $\bar{s}$,  and target structure, $\bar{\Gamma}$, 
$n(i,m)$ is the number of amino acids of type $i$ in the environment 
$m$ and $\epsilon(i,m)$ is the score associated with it (10). 
For a given amino acid $i$, each of the $\epsilon(i,m)$'s may be shifted
by the same arbitrary constant so that, without loss of generality,
one may set $\sum _{m} \epsilon (i,m) =0$.
The advantages of such an environmental scoring
function over pair-wise interactions between amino acids
are its simplicity and the far greater ease of incorporating gaps
in both sequence and in structure. 

Our focus is on determining the score quantifying the match of
a sequence to a putative native state structure,
the most common approach for which  utilizes statistical 
considerations (11-13),
based on counting the number of amino acids in a given 
environment in the native state. 
Pioneering work by Bowie et al. (10)
has shown that a simple statistically based
approach with an environmental score leads to excellent results for the
inverse folding problem. 

Our studies used a training set of 387 proteins 
(see Table I in Supplementary Information)
from the PDBselect (6,14)
consisting of sequences varying in length from 44 to 1017
with low sequence homology and covering many different 3D-folds
according to the
SCOP classification (15).
Additional criteria used in selecting the
proteins in the training set were: a) the protein structure
was obtained through X-ray crystallography, b) the structures were monomeric,
c) the determined structures missed no more than two amino acids.
The same criteria were used to obtain a test set of 213 distinct proteins
(Table 2 in Supplementary Information) with lengths ranging between 54
and 869. For each structure, we
used a simple environmental classification which consists
of the local secondary structure ($\alpha$-helix, $\beta$-strand or other)
and the exposed area evaluated as the ratio between the accessible
area of each amino acid, X, of its native sequence
(having this structure as its native state) and the corresponding area
in a Gly-X-Gly 
%CHANGE_LEFT
extended 
%CHANGE_RIGHT
structure. The values of the exposed area were
divided into three categories of small, medium and large exposures
corresponding to $< 10 \%$, $10 - 50\%$, and $> 50\%$ respectively. 
Thus the scoring function consists of nine parameters
for each amino acid corresponding to each of the nine environments that
it might be found in.

\vspace*{0.5cm}
%\noindent
%\newpage
\noindent
{\Large {\bf Materials and Methods}}

We begin by applying  the ideas of
Bowie et al. (10) to the threading problem. 
The statistical score $\epsilon _s(i,m) $ associated with amino acid $i$
in an environment $m$ is readily deduced using the expression
\begin{equation}
\epsilon _s(i,m)\;=\;-\ln \left[ P(i,m) / P(i) \right] \;\;,
\end{equation}
where $P(i,m)$ is the probability of finding an amino acid of type $i$ 
in the environment of type $m$ and $P(i)$ is the probability of finding 
an amino acid of type $i$ in any environment.   
Both $P(i,m)$ and $P(i)$ are determined from a knowledge of the sequences
and native state structures of the proteins in our training set. In order
to assess the quality of the extracted scores, we carried out threading tests
on all but the largest protein in the training set itself. Each protein
sequence was mounted on its own native state structure and on every fragment
(of the correct length chosen without insertions and deletions) 
of all the larger proteins. The exposed area for the amino acid mounted
on a fragment was assumed to be the same as that in the whole protein
from which the fragment was extracted. 
As we shall see later, this may be a poor approximation
when the size of the fragment is much smaller than the whole protein.
In each case, Eqn. (1) was used to determine the scoring function.
While the technique is simple, 
the results of gapless threading tests are only moderate 
-- the native state structure
is correctly recognized for $69\%$ of the proteins.  
In a recent paper, Baud and Karlin (16) considered 418
proteins and determined the frequencies of occurrence of the twenty
amino acids in nine environments which were defined in a way similar
to ours. We have converted their frequencies into statistical scores
(which turn out to be similar to the statistical scores derived from
our training set), using
equation (2), and find 54 failures in our set of 213 proteins.
This moderate performance may be due to the fact that
the form of the scoring function is too simple.
% or the statistical approach is flawed. 
Support for this comes from 
earlier work which has pointed out the difficulty of determining the optimal 
interactions that stabilize the native state of even 1 protein (crambin) 
with a more complex scoring  
function involving 210 pairwise interactions (17). 
An alternative possibility, that the statistical approach is
flawed (18)
would be of more serious concern because such statistical
schemes are commonly used in the protein folding problem.

We turn to a demonstration that an alternative
strategy based on ideas originating in statistical physics and neural networks
provides a powerful framework for tackling the threading problem.
Following the pioneering work of Friedrichs and Wolynes (19) and especially
Goldstein et al. (20), 
the basic idea is to postulate the form of a scoring function  
and to choose its parameters to ensure
that the true native states of proteins with known structures 
(learning set) correspond to better (lower) scores
than when the sequences are housed in competing decoy  
conformations (17-28).
An important advantage
of this procedure is that it can be used to verify whether the
chosen form of the scoring function is equal to the task or not. 
Indeed, one may start with the simplest form of the scoring function and 
systematically expand the parameter space until the optimal interactions
are learned. 
The statistical procedure considers proteins
and their native state structures, whereas the learning procedure has
information on competing structures as well.
%Even though some amino acids may statistically occur more often
%than other they can have lower magnitudes of the learned scores
%and thus overall contribute less to the score of a protein.
%CHANGE_LEFT
Our scheme is similar in spirit to that of previous work with
the important differences that we consider an
environmental scoring function instead of a pairwise contact potential
and we optimize the energy gap without any normalization.
%CHANGE_RIGHT

%The total number of inequalities (obtained by considering, for each sequence 
%in the training data set, all pieces of the native state structures of longer
%proteins in the training set as decoys) 
%is over 13 million making the problem technically difficult.
The total number of inequalities
(one obtains the inequalities for each sequence in the training data set
by considering as decoys all pieces of the native state structures of longer
proteins in the training set)
is over 13 million making the problem technically difficult.
For a given protein, each decoy leads to a linear inequality
of the form  
$\sum_{i=1}^{20} \sum_{m=1}^{9} \left[ n(i,m)^{D} - n(i,m) \right] 
\epsilon(i,m) > 0$, where $n(i,m)^{D}$ is the number of amino acids of 
type $i$ found in the environment $m$ in the given decoy. 
The perceptron procedure is a simple technique based on neural networks 
for simultaneously solving a 
set of such linear inequalities (29). 
We used this procedure to optimally choose the 180 parameters in order
to ensure that the worst inequality
(among the more than 13 millions) was satisfied as well as possible
and that threading tests on the training set were 100\% successful.
%Note that the sum of entries in each row has been chosen to be zero and 
%the sum of the squares of all entries has been set to $180$. 

\vspace*{0.5cm}
%\newpage
\noindent
{\Large {\bf Results}}

We describe the results of several tests and a biological 
interpretation of these parameters: 

\vspace*{0.5cm}

\noindent {\bf Learning procedure versus statistical approach:}~

Figure 1 shows a plot of the parameters determined using the
statistical approach versus those deduced by the learning procedure. 
The poor correlation 
is consistent with the qualitatively different performance levels 
in threading. It underscores the 
fundamental difficulties of the statistical approach and points to the
advantage of learning the optimized parameters in a systematically
expanded parameter set.  

\vspace*{0.5cm}

\noindent {\bf Threading tests:}~
%The native state of crambin (which was not part of the training set)
%is recognized in threading. This result is encouraging because of earlier 
%difficulties in learning pairwise parameters for this protein$^{17}$. 
%Possible difficulties may be the improper choice of decoy conformations
%or the form of the scoring function. In addition, tests on a new set of 
%proteins (1coa, 2ait, 1fkb, 1pks, 3chy, 1aps, 
%1fnf, 1srl, 1csp, 1c9o) also showed complete success in threading.

The couplings $\epsilon _{387}(i,m)$ obtained based on learning the 
native states of the 387 proteins in the training set were subjected to
threading  on the test set containing 213 distinct proteins 
(Table II in Supplementary Information) and the decoys obtained from their
native state structures.  In contrast to the performance of the statistical
parameters, for which one is unable to correctly recognize the native states
of 76 of the 213 proteins, the number of failures when one uses the learned
parameters is 23.  The failed proteins are modest in size and have sequence
lengths ranging between 54 and 131.  
The ranks of the native states, defined as the number of better performing
decoys, of the failed proteins are plotted in Figure 2 as a function
of the sequence length for both sets of parameters (note the dramatically
different scales of the $y$-axis). For the poorest performer,
using the perceptron based method, there are 102
decoys (out of 37617) that perform better than the native state
(protein 1abq of length 56) while the corresponding    %kinase
number for the statistically derived parameters is 29424 
(out of 31436 decoys for protein 1vqb of length 86).
%DNA binding protein
We have checked that around half of the failures
are spurious  for the case of the learned parameters
and arise because the exposed areas for the winning decoy,
which is a piece of the
native state structure of a longer protein, is quite different from that 
determined for the whole protein.  
This  effect of an inaccurate assignment of the exposed area is strong
only for small sized proteins. The remaining failures (a total of
5 \%) is likely due to the identification of genuine competitors to
the native state or because the winning decoy is not 
a viable structure for the sequence under consideration (9).

We also tested the $\epsilon _{387}$ parameters on all 600 proteins 
(Table I and II in Supplementary Information) and 
the decoys obtained using all 600 native state structures.
There were 57 failures whereas the statistically derived parameters resulted
in 209 failures.
We used the perceptron procedure (29)
to learn the scoring parameters in order to
ensure that the native state of all the proteins in the training and test set 
were recognized with 100 \% success and the energies of all
decoys were pushed up as much as possible compared to the native state energies.
In the rest of the paper, we will use this refined set of optimal 
parameters $\epsilon (i,m)$ (Table I) to carry out our further studies.
Note that the sum of the first nine entries of each row in Table I
is equal to zero and the sum
of the squares of all such 180 entries has been chosen to be 180.

The native state of crambin (1crn) which was not part of the training set,
%but a protein 1cbn, which has one mutation
%compared to 1crn, was in the set of the 387 proteins)
is recognized in threading. This result is encouraging because of earlier 
difficulties in learning pairwise parameters for this protein (19). 
It should be noted, however, that a single amino acid mutation of 1crn, the
protein 1cbn, was present in the basic learning set of 387 proteins.

%CHANGE_LEFT
As a further test, we selected 26 Globin proteins from the RCSB website
(http://www.rcsb.org/pdb/) which were in the DEOXY form, which were
not mutated and whose structures are resolved well.
Strikingly, 23 of the 26 proteins correctly picked their own native state
from among the millions of decoy conformations obtained from the fragments of
the 600 proteins in the training and test sets described previously.
For the 3 other cases, fragments from the Globin family were picked to be
the best structure.
Indeed, the scores of the Globin proteins on fragments of
other Globin structures were generally lower than on fragments of structures
of unrelated proteins underscoring the quality of our scoring function.
%CHANGE_RIGHT
\vspace*{0.5cm}

\noindent {\bf Biological interpretation of learned parameters:}~ 

Let us begin with a geometrical picture of $\epsilon (i,m)$, 
considered to be twenty
vectors of nine components each.
For a given amino acid $i$, the components of
the nine-dimensional vector,
labelled by the index $m$, capture the propensities of that
amino acid to be in each of nine environments. 
Each environment may be thought of as representing
an axis in an orthogonal 9-dimensional space.
Singular value decomposition (28, 31) %(SVD) 
affords a simple
prescription for dimensional reduction by the optimal choice of a new
set of orthogonal axes.  In this new reference frame, the original 
vectors span a lower-dimensional space and the axes may be conveniently
rank-ordered in importance.

The SVD theorem (31) states that the $20 \times 9$ 
%CHANGE_LEFT
(non-square) 
%CHANGE_RIGHT
matrix 
$\epsilon$
can be written as 
%%%%%%$$ \epsilon\;=\;U \Sigma V^T, $$.
\begin{equation}
$$ \epsilon\;=\;Y V^T, $$
\end{equation}
where $Y$ is a 20$\times $9 dimensional matrix and $V$ is a 9$\times $9
dimensional matrix. The superscript $^T$ denotes the transpose matrix.
The matrix $Y$ is given by $Y\;=\; U \Sigma$, where $\Sigma$ is
a $20\times 9$ dimensional matrix whose elements
%the matrix elements of $\Sigma$ 
are all zero except for the
diagonal terms, $\Sigma _{n,n} $, $n$=1,...,9. %which are 
These diagonal terms are equal to the
the square roots, $\sigma _n$, of the common eigenvalues
of $\epsilon \epsilon ^T$ and $\epsilon ^T \epsilon $.
The $\sigma _n$'s are called singular values
and are assumed to be rank ordered so that $\sigma _1$ is the largest.
Here, they are: 10.59, 5.02, 3.98, 3.42, 2.57, 2.09, 1.77, 0.95, and 0.0.
The columns of $V$, denoted
by $V_k$, are the eigenvectors corresponding to the rank ordered eigenvalues
of the matrix $\epsilon ^T \epsilon $ 
and the columns of the $20 \times 20$ matrix
$U$, denoted as $U_k$, $k$=1,...,20, are
determined by the formula $U_k=\frac{1}{\sigma _k} \epsilon V_k$ 
(when the singular values are non-zero; the other cases are irrelevant
for the reconstruction of the $\epsilon $ parameters).
%The columns of $U$ are the eigenvectors of $\epsilon \epsilon ^T$.
The result of the SVD transformation is that $\epsilon(i,m)$ may now be
represented  as a sum of contributions that diminish in an overall
sense as one considers smaller singular values.
The $n$'th contribution is given by $y_{(n)}(i) \; v^T_{(n)}(m)$,
where the first factor depends on the amino acid and the second
on the environmental index. Thus $v^T_{(n)}$ are the new orthogonal
and normalized directions --- or modes --- in the space of environments.

Figure 3 shows the three most dominant contributions, corresponding to the top
three singular eigenvalues. Each contribution is displayed in two panels. 
The upper panels show the mode as a function of the nine environmental
parameters. The lower panels show the corresponding amplitudes $y _{(n)}$ 
plotted so that $y _{(n)}$ increases monotonically.

The first mode is dominant for 13 amino acids: C, F, I, V, H,
S, T, N, P, Q, E, R, and K. The second mode
is the leader for W, M, Y, and D and the third for L and A.
The remaining amino acid, G, is dominated by the
fifth mode.
The first mode provides the overall dominant behavior and
strongly distinguishes between the buried and exposed 
environments in a monotonic way regardless of 
the secondary structure
--- $y _{(1)}$ allows one to arrange
the amino acids into buried and exposed groups depending on whether
it is large and positive or large and negative.
One may further subdivide the two basic groups of 
buried (B) and exposed (E) amino acid into subgroups:
B$_1$, B$_2$, B$_3$, E$_1$, and E$_2$.
The division is illustrated in Figure 3
and corresponds to occurrences of more rapid variations in
$y _{(1)}$ as one moves from one amino acid to the next.
The key point is that the amino acids in B$_1$ 
have a strong tendency to be buried and the charged amino acid K
in E$_2$  has a strong tendency to be exposed, and
most of the amino acids are more sensitive to the degree of
burial than to other considerations.
This tendency for burial is usually associated with
hydrophobicity in the
protein folding problem (32-34).
The hydrophobic amino acids F, I, V, L, and A do 
belong to group B but this group 
also contains polar amino acids. Cysteine, C, shows
the strongest propensity to be buried. It should be noted that
a pair of C's may form a strong contact by establishing a disulfide bridge.
Of the 896 C-C contacts generated in our study of 600 proteins,
402 had both C's buried whereas only in four cases
were both of the C's exposed (independent of the secondary structure).
This tendency alone yieds a high statistical score
for C being buried. (Note that 37\% of the structural sites
of the 600 proteins are classified as buried, 40 \% as medium, 
and 33 \% as exposed).
The learned score is even further accentuated
because most of the decoys correspond to C being not buried
and  stability of the native state with respect to decoys is
enhance by such an adjustment.

%We have considered 81 cases of forming such a contact, depending
%on what kind of the nine environments each C is in. In our study of
%the full set of 600 proteins, 
%the dominant occurrence and the lowest score 
%were associated with both C's being embedded regardless of the 
%nature of the secondary structure. The corresponding values were
%two orders of magnitude larger than in the exposed-exposed case.
%For instance, among the all 896 C-C contacts
%generated, 402 were of the embedded-embedded type and 4
%were exposed-exposed.

%and it is done in a way explained in the context of the pair wise
%Miyazawa-Jernigan couplings in ref. \cite{Cieplak}. 
%Basically, the divisions correspond to occurrences of more rapid
%variations in $y _{1)}$ along the amino acidic axis.
%The division into the five groups through the
%environment-based couplings is different than in the
%Miyazawa-Jernigan case. However, the overall split into the
%hydrophobic and polar amino acids is essentially the same,
%except for the interchange of G and H.
%Hydrophobicity emerges as a key factor determinig the properties
%of the amino acids since it underlies the interpretation of the first mode.

The remaining modes break the symmetry between the secondary 
structures. The second mode is neutral to $\alpha$ and favors
(disfavors) $\beta$ (loop) when the coefficient $y _{(2)}$ is negative.
It shows a strong preference
for amino acids, such as W, with a large negative $y _{(2)}$
to be in a $\beta$-strand with a large exposed area and for
amino acids, such as D, with a large positive $y _{(2)}$
to be in loops with a large exposed area.
The third mode introduces a preference for C, F, K, etc. to be
in $\beta$-strands with medium exposure and for L, P, and A, etc.
to stay either in exposed $\beta$-strands or in buried loops and avoid
exposed helices.

%We have considered alternative global characterizations of amino acids.
%One of them is the propensity
%of an amino acid to be in a given secondary structures. Another is
%the preference for a given kind of environment. For instance,
%$\epsilon (i,1) + \epsilon (i,2) + \epsilon (i,3) $ is a measure
%of propensity of the $i$'th amino acid to be in an $\alpha$ helix and 
%$\epsilon (i,1) + \epsilon (i,4) + \epsilon (i,7)$ characterizes the
%`willingness' to stay buried. This global measure of the buriedness,
%when ranked ordered, turns out to be alligning the amino acids according to
%CFIWVLMAGYSHDTNPQERK. This order generally agrees with the one provided
%by the buriedness index of the first SVD mode.
%Furthermore, the global propensity measures also support the
%interpretation of the second and third modes.

\vspace*{0.5cm}

\noindent {\bf Protein design:}~ 
We turn now to an extension of our studies to protein design or the
inverse folding problem. In analogy with equilibrium statistical mechanics,
the probability that a sequence $\bar{s}$ is housed in a structure $\bar{\Gamma}$
is given by (35-38)

%\begin{eqnarray*}
\begin{eqnarray}
P_{\bar{\Gamma}}(\bar{s})  =  {e^{-S(\bar{s},\bar{\Gamma})/T} \over \sum_{\Gamma^{\prime}} e^{-S(\bar{s},\Gamma^{\prime})/T}} \equiv  {e^{-S(\bar{s},\bar{\Gamma})/T} \over e^{-F(\bar{s})/T}} 
%\end{eqnarray*}
\end{eqnarray}
where $T$, here, is a fictitious temperature, the score $S$ has been assumed to
play a role analogous to the energy and $F$, the free score, plays the role of 
the free energy. The key point is that in the limit of $T \rightarrow 0$  and when
$\bar{\Gamma}$ is the native state structure of $\bar{s}$, $P \rightarrow 1$.
In this limit, therefore, the $\verb|"|$free score$\verb|"|$ which is a function
of the sequence alone approaches the score of the sequence in its native state. 
The last column of Table 1 shows the 
average contribution to the native state scores, $S_i$, 
from each type of amino acid in the
various environments. It is defined by 
%CHANGE_LEFT
$S_i\;=\;\frac{1}{N_i} \sum _{k=1}^{N_i} \epsilon(i,m(k))$,
%CHANGE_RIGHT
% $S_i\;=\;\frac{1}{N_i} \sum _{m=1}^{N_i} \epsilon(i,m)$,
where the sum is over the $N_i$ occurrences of amino acid $i$
in the native state of all 600 proteins in the training and test sets.
The zero $\verb|"|$temperature$\verb|"|$ free score of a sequence may then be
readily deduced without any knowledge of the structure, by adding up these
contributions for the amino acids in the sequence. 
Figure 4 shows a plot of the native state score
versus the sequence free score for all 600 proteins. 
The latter, which has no structure dependent
information, provides a reasonable approximation to the actual 
native state score.
We have verified that both are linearly proportional to the 
protein length and for the longer proteins,
the native state score is somewhat higher than the free score due to  
the increasing tendency toward frustration as the sequence length increases.
For design purposes, the free score provides a measure of the score one is
entitled to expect in a typical native state
structure and the lower the score in the target
native state structure with reference to the free score,
the better is the design. 

\vspace*{0.5cm}

\noindent {\bf Stability of cold shock proteins:}~
We used the learned parameters to provide a molecular
interpretation of the different thermal stabilities of a pair of cold
shock proteins (39), one of which is mesophilic Bacillus subtilis 
(Bs-CspB: 1csp) and the other thermophilic Bacillus caldolyticus (Bc-Csp: 1c9o).
The former has a score of -34.80 in the native state,
whereas the latter is more stable with a score of -41.64. 
More strikingly, the free scores are -28.64 and -27.97 respectively underscoring
the much better design of the thermophilic protein.  
We also used the conformation space of all decoys to estimate 
the $\verb|"|$heat capacity$\verb|"|$ of the 
two proteins as a function of temperature. 
The heat capacity which is a measure of the fluctuations in the score
(viewed as an energy) shows a peak as a function of the temperature 
in both cases. The peak
temperature, which is a measure of the folding transition temperature
of the protein,
%Strikingly,
%the ratio of the peak temperature in the two cases is found to be within
%$3\%$ of the experimentally observed values (35). 
%CHANGE_LEFT
of 1c9o is higher than that of 1csp, and
reflects the better thermal stability of 1c9o in accord with the experimentally
observed behavior (39). 
%CHANGE_RIGHT

\vspace*{0.5cm}
\noindent
{\Large {\bf Conclusion}}

In summary, we have shown that a straightforward learning scheme leads to the
determination of excellent environmental parameters which can be
used in simple threading tests. Our results point to the danger of
employing statistical procedures for estimating these values. 
The learned parameters capture information on the environments in
the competing structures in addition to that in the native
state structures and allows for a stabilization of the native state
with respect to decoy structures.
Our procedure validates the notion that in the simplest cases we have
studied here, a simple environmental scoring function is sufficient for
capturing the essential features of protein threading. Our method has the
distinct advantage of ease of expanding the parameter space and opens
up the possibility of using the scoring parameters determined here
as a starting point for
learning the penalty parameters characterizing insertion and deletion.

This work was supported by grants from NASA, INFM and MURST (Italy), 
the Donors of the Petroleum Research
Fund administered by the American Chemical Society, PNU research fund and 
KBN (grant number 2P03B-146-18).

%------------------------------------------------------
%------------------------------------------------------
\newpage

\begin{enumerate}

\item Anfinsen, C. (1973) 
%Principles that govern the folding of protein chains. (1973)
{\it Science} {\bf 181}, 223-230.

\item Wolynes, P. G., Onuchic, J. N., \& Thirumalai, D. (1995)
%Navigating the folding routes.
{\it Science} {\bf 267}, 1619-1620.

\item Dill, K. A. \& Chan, H. S. (1997)
%From Levinthal to pathways to funnels.
{\it Nature Struct. Biol.} {\bf 4} 10-19.

\item Fersht, A. P. (1998)
{\it Structure and mechanism in protein science: A guide
to enzyme catalysis and protein folding}, New York, Freeman.

\item Baker, D. A. (2000)
%A surprising simplicity to protein folding.
{\it Nature} {\bf 405}, 39-42.

\item Berman, H. M., Westbrook, J., Feng, Z., Gilliland, G.,
Bhat, T. N.,  Weissig, H., Shindyalov, I. N. \& Bourne, P.E. (2000)
%The Protein Data Bank.
{\it Nucl. Acid. Res.} {\bf 28}, 235-242.

\item Jones, D. T., Taylor, W. R., \& Thornton, J. M. (1992)
%A new approach to protein fold recognition.
{\it  Nature} (London) 
{\bf 358}, 86-89.

\item Chothia, C. (1992)
%One thousand folds for the molecular biologist.
{\it Nature} {\bf 357}, 543-544. 

\item
Ramachandran, G. N. \& Sasisekharan, V. (1968)
%Conformation of polypeptides and proteins.
{\it Adv. Prot. Chem.} {\bf 28}, 283-437.

\item Bowie, J., L\"uthy, R. \&  Eisenberg, D. (1991)
%A method to identify protein sequences that fold into a known
%three-dimensional structure.
{\it Science} {\bf 253},
164-170.

\item Tanaka, S. \&  Scheraga, H. A.  (1976)
%Medium- and long-range interaction parameters between amino acids
%for predicting three dimensional structures of proteins.
{\it Macromolecules} {\bf 9}, 945-950.

\item
S. Miyazawa, S. \& Jernigan,  R. L. (1985)
{\it Macromolecules} {\bf 18}, 534-552. 

\item
Zhang, C. \& Kim, S. (2000)
%Environment-dependent residue contact energies for proteins.
{\it Proc. Natl. Acad. Sci.} {\bf 97}, 2550-2555.

\item Hobohm, U. \& Sander, C. (1994)
%Enlarged representative set of protein structures.
{\it Prot. Sci.} {\bf 3}, 522-524.

\item Murzin, A. G., Brenner, S. E., Hubbard, T. \& Chothia, C. (1995)
%SCOP - a structural classification of proteins database for
%the investigation of sequences and structures.
{\it J. Mol. Biol.} {\bf 247}, 536-540.

\item Baud, F. \& Karlin, S. (1999)
%Measures of residue density in protein structures.
{\it Proc. Natl. Acad. Sci.} {\bf 96}, 
12494-12499.

\item Vendruscolo, M., Najmanovich, R. \& Domany, E. (1999)
%Protein folding in contact map space.
{\it Phys. Rev. Lett.} {\bf 82}, 656-659. 

\item Dima, R. I., Banavar, J. R. \& Maritan, A. (2000)
%Scoring functions in protein folding and design.
{\it Protein Sci.} {\bf 9}, 812-819.

\item Friedrichs, M. S. \& Wolynes, P. G. (1989)
%Toward protein tertiary structure recognition by means of associative
%memory Hamiltonians.
{\it Science} {\bf 246}, 371-373.

\item Goldstein, R., Luthey-Schulten, Z. A. \& Wolynes, P.G. (1992)
{\it Proc. Natl. Acad. Sci.} {\bf 89}, 9029-9033. 

\item Koretke K.K., Luthey-Schulten, Z. A. \& Wolynes, P.G. (1996)
{\it Protein Science} {\bf 5}, 1043-1059. 

\item
Maiorov, V. N. \&  Crippen, G. M. (1992)
%Contact potential that recognizes the correct folding of globular proteins.
{\it J. Mol. Biol.} {\bf 227}, 876-888.

\item
Mirny, L. A. \&  Shakhnovich, E. I. (1996)
%How to derive a protein folding potential? A new approach to an old problem.
{\it J. Mol. Biol.} {\bf 264}, 1164-1179.

\item
Clementi, C. Maritan, A. \& Banavar, J. R. (1998)
%Folding, design, and determination of interaction potentials
%using off-lattice dynamics of model heteropolymers.
{\it Phys. Rev. Lett.} {\bf 81}, 3287-3290.

\item
Dima, R. I., Settanni, G., Micheletti, C. Banavar, J. R. \&  
Maritan A. (2000)
%Extraction of interaction potentials between amino acids from native
%protein structures.
{\it J. Chem. Phys.} {\bf 112}, 9151-9166.

\item
Vendruscolo M., Mirny L.A., Shakhnovich E. I. \& Domany E. (2000)
{\it Proteins: Structure, Function, and Genetics} {\bf 41}, 192-201. 

\item
Tobi, D. \& Elber, R. (2000)
{\it Proteins: Structure, Function, and Genetics} {\bf 41}, 40-46. 

\item
Tobi, D., Shafran, G., Linial, N. \&  Elber, R. (2000)
%On the design and analysis of protein folding potentials.
{\it Proteins: Structure, Function, and Genetics} {\bf 40}, 71-85.

\item Krauth, W. \& Mezard, M. (1987)
%Learning algorithms with optimal stability in neural networks.
{\it J. Phys. A} {\bf 20}, 
L745-L752. 

\item Park, B. \& Levitt, M. (1996)
%Energy functions that discriminate X-ray and near-native folds
%from well-constructed decoys.
{\it J. Mol. Biol.}
{\bf 258}, 367-392. 

\item
Watkins, D. S. (1991) {\it Fundamentals of Matrix Computations}, Wiley, 
New York.

\item
Kauzmann, W. (1959)
%Some factors in the interpretation of protein denaturation.
{\it Adv. Protein Chem.} {\bf 14}, 1-63.

\item
Dill, K.A. (1990)
%Dominant forces in protein folding.
{\it Biochemistry} {\bf 29}, 7133-7155.

\item
Kamtekar, S., Schiffer, J. M., Xiong, H. Y., Babik, J. M. \& 
Hecht, M. H. (1993)
%Protein design by binary patterning of polar and nonpolar amino--acids.
{\it Science} {\bf 262}, 1680-1685.

\item Deutsch, J. M. \& Kurosky, T. (1996)
%New algorithm for protein design.
{\it Phys. Rev. Lett.}  {\bf 76}, 323-326.

\item
Seno, F., Vendruscolo, M., Maritan, A. \& Banavar, J. R. (1996)
%Optimal protein design procedure.
{\it Phys. Rev. Lett.} {\bf 77}, 1901-1904.

\item
Dima, R. I., Banavar, J. R.  Cieplak, M. \& Maritan, A. (1999)
%Statistical mechanics of protein-like heteropolymers.
{\it Proc. Natl. Acad. Sci.} {\bf 96}, 4904-4907.

\item
Micheletti, C., Maritan, A. \& Banavar, J. R. (1999)
%A comparative study of existing and new design techniques for
%protein models.
{\it J. Chem. Phys.} {\bf 110}, 9730-9738.

\item Perl, D., Mueller, U., Heinemann, U. \& Schmid, F. (2000)
%Two exposed amino acid residues confer thermostability on a cold shock
%protein.
{\it Nature Struct. Biol.} {\bf 7}, 380-383.

\end{enumerate}
%------------------------------------------------------

\newpage
\vskip 1.5cm

\noindent {\bf Table Captions} 

\noindent Table 1:~Table of $\epsilon$, the nine environmental scores for 
each amino acid. Large negative values indicate a strong preference for
the particular environment whereas large positive values indicate an aversion.
The last column shows $S_i$ which
is a measure of the average contribution 
of each amino acid to the native state score
and provides an estimate of the
expectation of the contribution of a given amino acid to the native state score.

\vfill
\eject

\centerline{\bf Table I. 180 Environmental Scores} 

\begin{tabbing}
xxxxxxxxxxxxxxxx\=xxxxxxx\=xxxxxxx\=xxxxxxx\=xxxxxxx\=xxxxxxx\=xxxxxxx\=xxxxxxx\=xxxxxxx\=xxxxxxx\=xxxxxxx \kill
               \> \> {\bf $\alpha$ } \> \> \> {\bf $\beta$ } \> \> \> {\bf Other} \> \> {\bf S$_i$}\\
               \> \underline{~~~~~~~~~~~~} \> \underline{~~~~~~~~~~~~} \> \underline{~~~~~~} \> \underline{~~~~~~~~~~~~} \> \underline{~~~~~~~~~~~~} \> \underline{~~~~~~} \> \underline{~~~~~~~~~~~~} \> \underline{~~~~~~~~~~~~} \> \underline{~~~~~~} \> \underline{~~~~~~~} \\
{\it\bf Amino Acid} \> Small \> Med. \> Expo. \> Small \> Med. \> Expo. \> Small \> Med. \> Expo.  \\ 

{\bf CYS ~~(C)} 
\> -1.29 \> ~0.07 \> ~1.81 \> -1.78 \> -0.83 \> ~3.63 \> -1.24 \> -0.85 \> ~0.49 \> -1.06 \\
{\bf PHE ~~(F)} 
\> -0.90 \> -0.35 \> ~2.33 \> -1.77 \> -1.02 \> ~1.51 \> -0.26 \> -0.28 \> ~0.74 \> -0.73 \\
{\bf TRP ~~(W)} 
\> ~0.41 \> ~0.32 \> ~1.64 \> -1.18 \> -1.00 \> -1.02 \> ~0.57 \> ~0.50 \> ~0.91 \> -0.07 \\
{\bf ILE ~~~~(I)} 
\> -0.50 \> -0.27 \> ~0.38 \> -0.25 \> -0.39 \> ~0.61 \> -1.05 \> ~0.56 \> ~0.92 \> -0.29 \\
{\bf VAL ~~~(V)} 
\> ~0.42 \> ~0.06 \> -0.12 \> -1.48 \> -0.64 \> ~0.89 \> -0.28 \> ~0.57 \> ~0.58 \> -0.31 \\
{\bf MET ~~(M)} 
\> -0.26 \> -0.36 \> ~0.65 \> -0.52 \> ~0.71 \> ~1.26 \> -0.24 \> -0.77 \> -0.47 \> -0.30 \\
{\bf LEU ~~~(L)} 
\> -0.33 \> -0.16 \> ~0.09 \> -0.32 \> ~0.83 \> -0.76 \> -0.54 \> ~0.77 \> ~0.41 \> -0.10 \\
{\bf GLY ~~~(G)} 
\> ~0.36 \> ~1.16 \> ~0.73 \> -0.05 \> ~0.16 \> ~0.14 \> -0.49 \> -0.95 \> -1.06 \> -0.48 \\
{\bf TYR ~~~(Y)} 
\> ~0.13 \> ~0.83 \> -0.06 \> -0.42 \> -1.18 \> -0.23 \> ~0.23 \> ~0.08 \> ~0.63 \> ~0.00 \\
{\bf ALA ~~~(A)} 
\> -0.40 \> -0.05 \> -0.13 \> ~0.27 \> ~0.50 \> -0.15 \> -0.23 \> ~0.35 \> -0.25 \> -0.06\\
{\bf HIS ~~~~(H)} 
\> ~1.05 \> -0.60 \> -0.82 \> ~0.62 \> ~0.56 \> ~0.14 \> -0.29 \> -0.08 \> -0.57 \> -0.09 \\
{\bf ASP ~~~(D)} 
\> -0.29 \> -0.79 \> -0.90 \> ~1.31 \> ~0.93 \> ~1.32 \> ~0.59 \> -0.99 \> -1.17 \> -0.60 \\
{\bf SER ~~~~(S)} 
\> -0.31 \> -0.01 \> -0.98 \> ~0.48 \> ~0.78 \> -0.75 \> ~1.00 \> -0.32 \> -0.10 \> ~0.03 \\
{\bf THR ~~~(T)} 
\> ~0.80 \> ~0.49 \> -0.46 \> ~0.55 \> -0.50 \> -0.80 \> ~0.74 \> -0.34 \> -0.48 \> ~0.01 \\
{\bf ASN ~~~(N)} 
\> ~0.67 \> -0.66 \> -0.66 \> ~1.34 \> ~0.60 \> -0.06 \> ~0.55 \> -0.48 \> -1.30 \> -0.39 \\
{\bf PRO ~~~(P)} 
\> ~2.35 \> -0.28 \> -0.88 \> ~1.32 \> ~1.03 \> -0.30 \> -1.02 \> -0.62 \> -1.61 \> -0.65 \\
{\bf GLN ~~~(Q)} 
\> ~1.74 \> -0.84 \> -1.24 \> ~0.94 \> -0.87 \> -1.07 \> ~1.32 \> ~0.01 \> ~0.01 \> -0.26 \\
{\bf GLU ~~~(E)} 
\> ~0.83 \> -0.81 \> -1.28 \> ~1.67 \> -0.21 \> -0.67 \> ~1.60 \> ~0.04 \> -1.16 \> -0.53 \\
{\bf ARG ~~~(R)} 
\> ~2.29 \> -0.80 \> -1.37 \> ~1.37 \> -1.16 \> -1.35 \> ~1.82 \> ~0.13 \> -0.94 \> -0.38 \\
{\bf LYS ~~~~(K)} 
\> ~1.20 \> -1.13 \> -1.77 \> ~4.32 \> -1.43 \> -1.91 \> ~2.38 \> -0.32 \> -1.35 \> -1.11 \\

\end{tabbing}

\vfill
\eject

%\vspace*{-3cm}
%\vspace*{-2.6cm}
\vspace*{4.5cm}

\centerline{\bf Supplementary Table I. PDB codes for 387 learning proteins} 

{\tiny
\begin{tabbing}
xxxxxx\=xxxxxx\=xxxxxx\=xxxxxx\=xxxxxx\=xxxxxx\=xxxxxx\=xxxxxx\=xxxxxx\=xxxxxx\=xxxxxx\=xxxxxx\=xxxxxx\=xxxxxx \kill

1CII\>1KCW\>4HB1\>1DHX\>1BLE\>1SQC\>1AB4\>1PKP\>1FIY\>1RGS\>1TDJ\>1FSZ\>8LDH\>6ICD\\
1AEP\>1ANV\>5PTD\>1A0I\>1B6E\>1SIG\>1DIV\>1KXU\>1BVB\>1BAJ\>1A6F\>1LRV\>2ITG\>1CBY\\
1LXA\>1914\>1CC5\>1A0P\>1OHK\>1JON\>1PJR\>1A7J\>1AUA\>1GLN\>1IHP\>1HLB\>1ZAP\>2STV\\
1A17\>1RDR\>1RLW\>1UBY\>2SAS\>1BCO\>1A8Y\>1AX8\>2LIV\>1AN8\>2OMF\>1A41\>1C25\>1AK5\\
1AQT\>1AJ6\>2FXB\>1BOB\>1INP\>1CYX\>1XSM\>1BIA\>1CPT\>1OPR\>1PLQ\>1AUQ\>1KIT\>1CTN\\
1DHR\>1OBR\>1RCB\>1A26\>1CIY\>1GPC\>1PFO\>1GRJ\>1BY9\>1MAZ\>1LBA\>1KTE\>1AM2\>1BB9\\
1HTP\>1BIX\>1TUL\>1DRW\>1AQE\>2GSQ\>1DHS\>5EAU\>1CFR\>1GEN\>1BR9\>1ACC\>1YGS\>1B5L\\
1IAM\>1A32\>1RMD\>1PEA\>1SEK\>1KLO\>1OXA\>1CRB\>2TCT\>1ESC\>1TFR\>2NG1\>1LCI\>1PHT\\
1ALY\>1VIN\>1A6Q\>1A76\>1A1X\>1CSN\>1TIG\>1A8H\>1BTN\>1CDY\>1CFB\>1MSC\>1AMX\>1HOE\\
1UOX\>2PGD\>1BV1\>2PLC\>4MT2\>1SRA\>1DDT\>1NSJ\>1UOK\>1POC\>1SUR\>1GOX\>1GSA\>1MJC\\
2PIA\>1LKI\>1BY2\>1SKF\>1BIF\>1PBV\>1ALO\>1RMG\>2I1B\>1DPE\>1AJ2\>4PAH\>1FCE\>1PNE\\
1BF2\>1AZ9\>1A53\>3TDT\>7TAA\>1OPC\>1PTQ\>1BEA\>1PUC\>1FUA\>1RSS\>1ECL\>1SKZ\>1NEU\\
1ALU\>1CUK\>1CA1\>1MAI\>1AD2\>1OPY\>1EDT\>1BHE\>1JDW\>1PHM\>1DXY\>1VOM\>1CEO\>1A8L\\
1TMY\>1SVB\>1AIL\>1WHO\>1JDC\>1SFP\>3TSS\>1DUN\>1IOW\>1PBE\>1GPR\>1A48\>4ENL\>2PII\\
3GCB\>1BG7\>1VLS\>1PUD\>2ABK\>1MDL\>1RKD\>1EUR\>1DMR\>1GND\>1UCH\>1BG2\>1AK0\>1UXY\\
2GAR\>1LCL\>1MML\>1POT\>1QNF\>1NPK\>1AYL\>1TIF\>1BD8\>1BDO\>1BG6\>1C3D\>1HYP\>2POR\\
1UAE\>1BJ7\>1TML\>1TYV\>1HCL\>2SAK\>1FNA\>1AL3\>2TGI\>2ACY\>1LST\>1LBU\>1AMP\>1NAR\\
1FAS\>2CBP\>1FMB\>1AXN\>1TUD\>1PDA\>1HA1\>1CV8\>1CHD\>1AMF\>1USH\>1CPQ\>1BM8\>1XWL\\
1BGC\>1AJJ\>1TFE\>1NKR\>1IDO\>1VJS\>1BHP\>1WAB\>1VIE\>1VHH\>1GCA\>1PDO\>1FDR\>1PMI\\
1SBP\>1GOF\>1AKO\>1MOF\>2GDM\>1FXD\>1FNC\>1GAI\>2HFT\>1OSA\>1VNS\>3CHY\>1ERV\>1DHN\\
1AQB\>1CNV\>119L\>1CEM\>1CXC\>1VCC\>1GVP\>2DRI\>1MBA\>1A3C\>1EDG\>1PHF\>16PK\>451C\\
1B6A\>1BKF\>1RZL\>5NUL\>1AOP\>1A8E\>1CVL\>1ARV\>1NIF\>3CYR\>1MRJ\>1ZIN\>1LAM\>1CSH\\
1KUH\>1PTF\>1BFG\>1BFD\>3PTE\>2AYH\>2MYR\>1NOX\>1AKR\>2A0B\>1A8D\>1MOQ\>1HFC\>1RA9\\
1TCA\>3GRS\>2CBA\>1KPF\>5ICB\>1AIE\>1KOE\>1WHI\>1RIE\>1MLA\>1HKA\>1OPD\>1FLP\>2MCM\\
1CYO\>1POA\>1BRT\>2HBG\>2SNS\>1XNB\>2RN2\>3SEB\>1BGF\>2END\>1YGE\>3VUB\>2CTC\>1HMT\\
1PPT\>1BQK\>1UTG\>1PLC\>1BK0\>1DCS\>1C52\>7RSA\>1OAA\>1MSI\>1YCC\>2PTH\>2SN3\>1AMM\\
1BX7\>1ATG\>2KNT\>1MUN\>1A7S\>1CTJ\>1BS9\>2IGD\>1NKD\>3SIL\>2ERL\>1A6M\>1CEX\>1IXH\\
1BYI\>1AHO\>1NLS\>2FDN\>3LZT\>1RB9\>3PYP\>1CBN\>1GCI\>    \>    \>    \>    \>    \\

\end{tabbing}
}

\vfill
\eject

\vspace*{2cm}

\centerline{\bf Supplementary Table II. PDB codes for 213 testing proteins} 

{\tiny
\begin{tabbing}
xxxxxx\=xxxxxx\=xxxxxx\=xxxxxx\=xxxxxx\=xxxxxx\=xxxxxx\=xxxxxx\=xxxxxx\=xxxxxx\=xxxxxx\=xxxxxx\=xxxxxx\=xxxxxx \kill

1AF5\>1FHE\>1CRY\>1ADT\>2CHR\>2LDX\>2UCZ\>1AVC\>1LDB\>1AR2\>1BMP\>1ABQ\>1TLK\>1ULA\\
1EIA\>1A1S\>1B5M\>1PEX\>3PHV\>1OJT\>2FHI\>6FIT\>1A43\>2TPT\>1YFM\>1LU1\>1CPY\>1HJP\\
1ILE\>1FBL\>1GWZ\>1BIK\>1PMT\>1A06\>4FXC\>1BAG\>1HUP\>1CYW\>1JSG\>1BMG\>1FSU\>2CND\\
2GPR\>1PBK\>2ABL\>1TMO\>1CYG\>2ALR\>1NAT\>1SBF\>2TDX\>1KAS\>1A6I\>4P2P\>1SZT\>1DOL\\
1FGS\>1AC5\>1HIB\>1QPG\>1CQA\>1DOT\>1MKP\>1ANN\>8OHM\>1ASS\>1JNK\>1BET\>8CHO\>2ASR\\
1AD6\>1BQG\>1DHY\>1A45\>1DIK\>1CIU\>3KAR\>1APA\>2PK4\>1HEY\>1BYT\>1ODD\>1ZXQ\>1FTS\\
1AA0\>1GDD\>1PHK\>1VIP\>1HAR\>1TSY\>1AW9\>1ENY\>1AVK\>1YVS\>1BFS\>5PNT\>1A0K\>1AIR\\
2DAP\>1ECY\>2ASI\>1ANU\>1CBG\>1FAJ\>1P38\>1A6O\>1BCG\>1A6L\>3ERK\>1BED\>1BKL\>1ENH\\
1BLU\>1A80\>1A8Z\>1KIV\>1MNC\>1A3K\>1XAA\>1A0B\>2PSR\>1B56\>1LPP\>1BK2\>3KVT\>1NFO\\
1INR\>1RGP\>1AQ1\>1TN3\>2E2C\>1GSH\>1A8P\>1AZ5\>1PVL\>1BKM\>1VPE\>1MHO\>1BK1\>1AE7\\
1FIL\>1MN1\>1RCI\>1RIS\>1HVF\>1ACF\>1BDB\>1AYI\>1NHP\>1IAG\>1CGT\>1AA2\>2EBN\>1BK9\\
1POH\>1ESL\>1HCZ\>1A58\>1AEW\>1BAM\>1SNP\>1AK2\>3GAR\>1EIF\>1CLC\>4RHN\>1DFX\>1A8B\\
1KVY\>1CYI\>1UKZ\>1DOI\>1BGP\>1TN4\>1CYJ\>1YMV\>1AX0\>1HFX\>1LRA\>1RCY\>1NUC\>1JUG\\
3PRN\>1ENP\>1LML\>1DYR\>1OBM\>1A44\>1ZRN\>2HTS\>1RFS\>1AMK\>1IVD\>1IFT\>1NFN\>1ARS\\
1AHR\>1XND\>1RDS\>1A68\>1VQB\>1ALQ\>1MRG\>1A7E\>2SLI\>1GBS\>3KIV\>1BZA\>1AT5\>2ERA\\
1A3D\>1GZI\>1A8S\>Ö\>    \>    \>    \>    \>    \>    \>    \>    \>    \>    \\

\end{tabbing}

}

\vfill
\eject

\vskip 1.5cm

\noindent {\bf Figure Caption} 

\noindent Figure 1: Plot of the optimal $\epsilon$ parameters %%shown in Table 1 
versus those determined using a statistical scheme, $\epsilon _s$, 
using a training set of 387 proteins.

\vspace*{0.5cm}

\noindent Figure 2: Results of the threading tests for 213 proteins arranged
according to their length, $N$. Only the failed cases are shown.
The top panel shows a plot of the number of decoys
that performed better than the native state structure versus $N$ whereas the 
bottom panel shows a similar plot for the couplings that
were determined statistically.
Note the disparity in the scales of the $y$-axes.

\vspace*{0.5cm}

\noindent Figure 3: The top three contributions to
$\epsilon (i,m)$ as emerging from the SVD analysis.
The numbers in the ovals indicate the mode number.
The letters at the top left of each segment of two panels
indicate amino acids (in the single letter code) for which
this particular mode is dominant.
The top panels in each segment show the modes -- the values of $v^T_{(n)}$
for the nine values of the environmental variable $m$. 
For each kind of secondary structure, the environments
are listed in order from the small to large exposure.
The bottom panels show the amino-acid-dependent weights $y_{(n)}$ with which
the displayed mode contributes to the score
in a given environment.

\vspace*{0.5cm}

\noindent Figure 4: Plot of the zero $\verb|"|$temperature$\verb|"|$ free score
and the native state score of each of the proteins 
in the training and test sets.  

\newpage

%FIGURE 1
\begin{figure}
\vspace*{1.5cm}
\epsfxsize=5.2in
\hspace*{-0.5cm}
\centerline{\epsffile{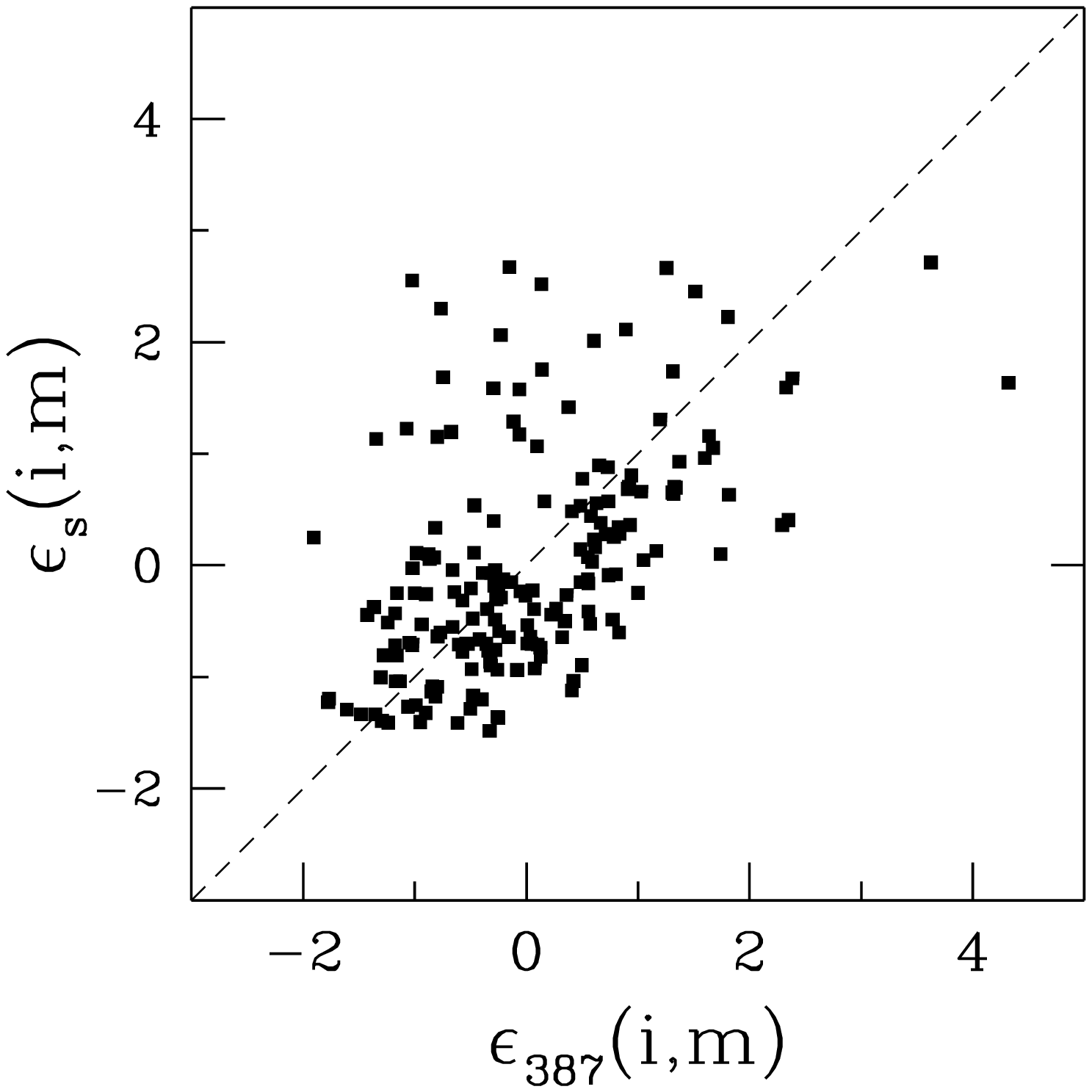}}
%\vspace*{1.8cm}
\vspace*{2.3cm}
\caption{ }
\end{figure}

%FIGURE 2
\begin{figure}
\vspace*{3cm}
\epsfxsize=5.2in
\hspace*{-0.5cm}
\centerline{\epsffile{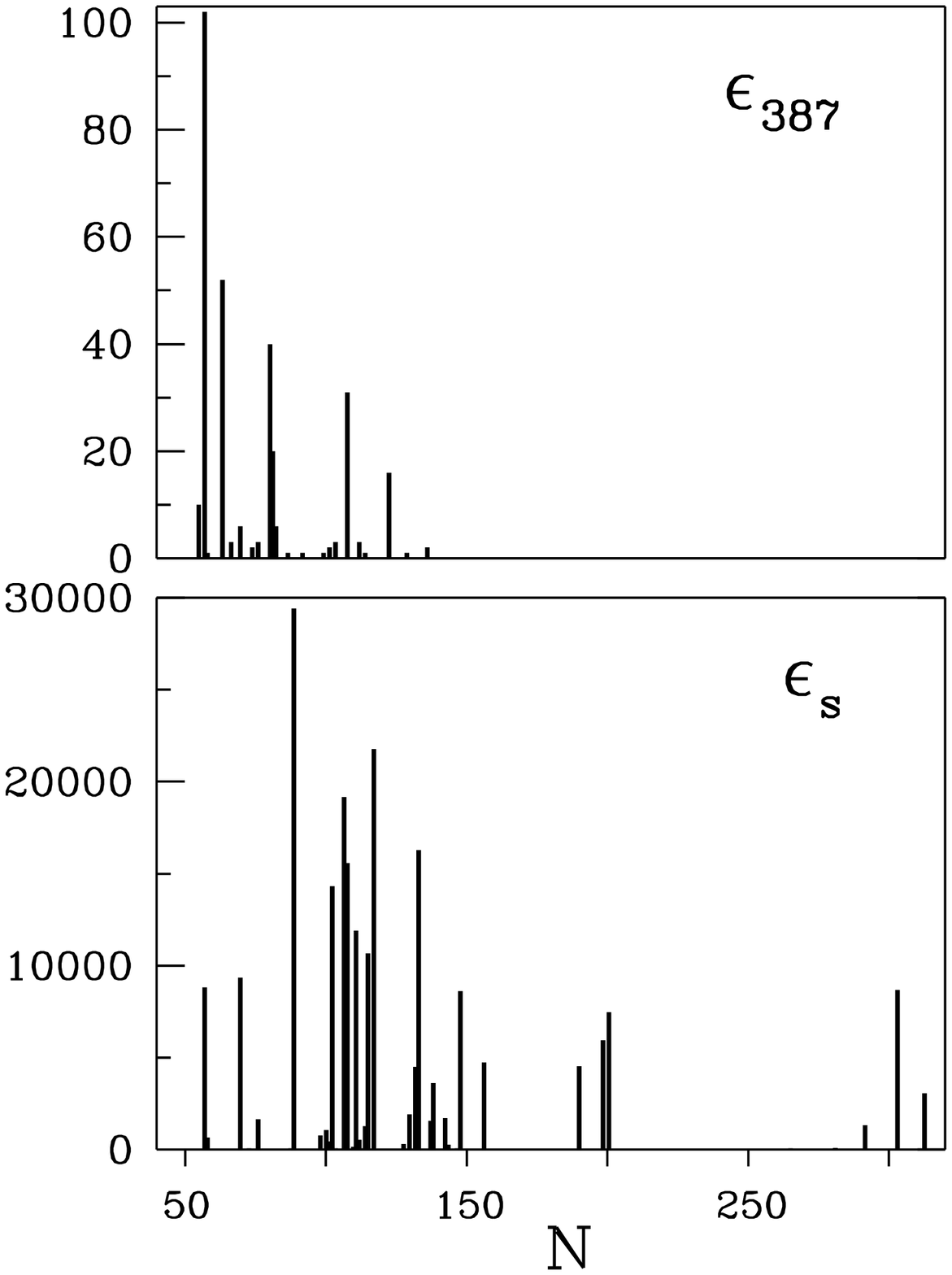}}
%\vspace*{1.8cm}
\vspace*{3.8cm}
\caption{ }
\end{figure}

%FIGURE 3
\begin{figure}
\vspace*{3cm}
\epsfxsize=5.2in
\hspace*{-0.5cm}
\centerline{\epsffile{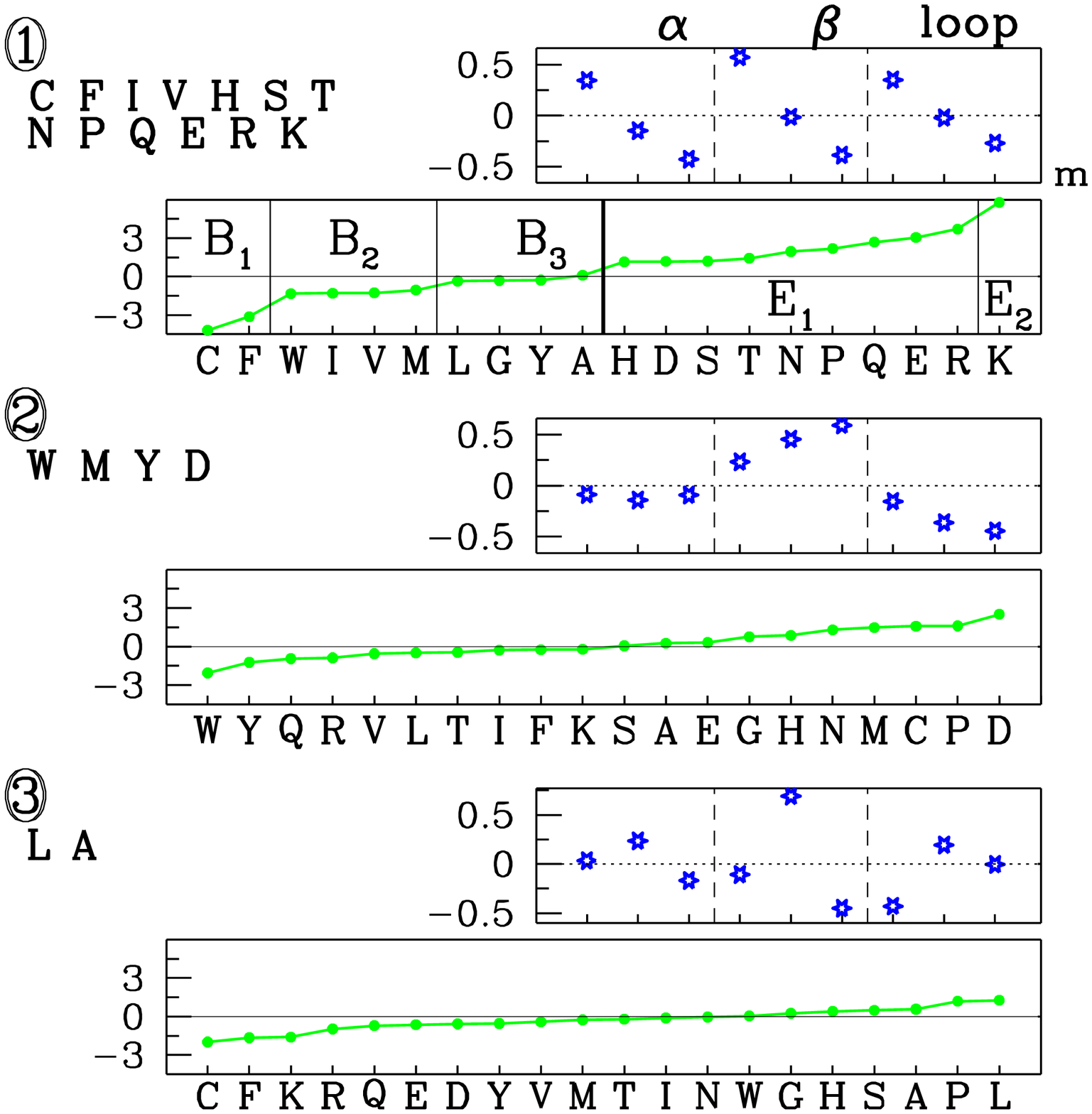}}
%\vspace*{1.8cm}
\vspace*{3.8cm}
\caption{ }
\end{figure}

%FIGURE 4
\begin{figure}
\vspace*{2cm}
\epsfxsize=5.2in
\centerline{\epsffile{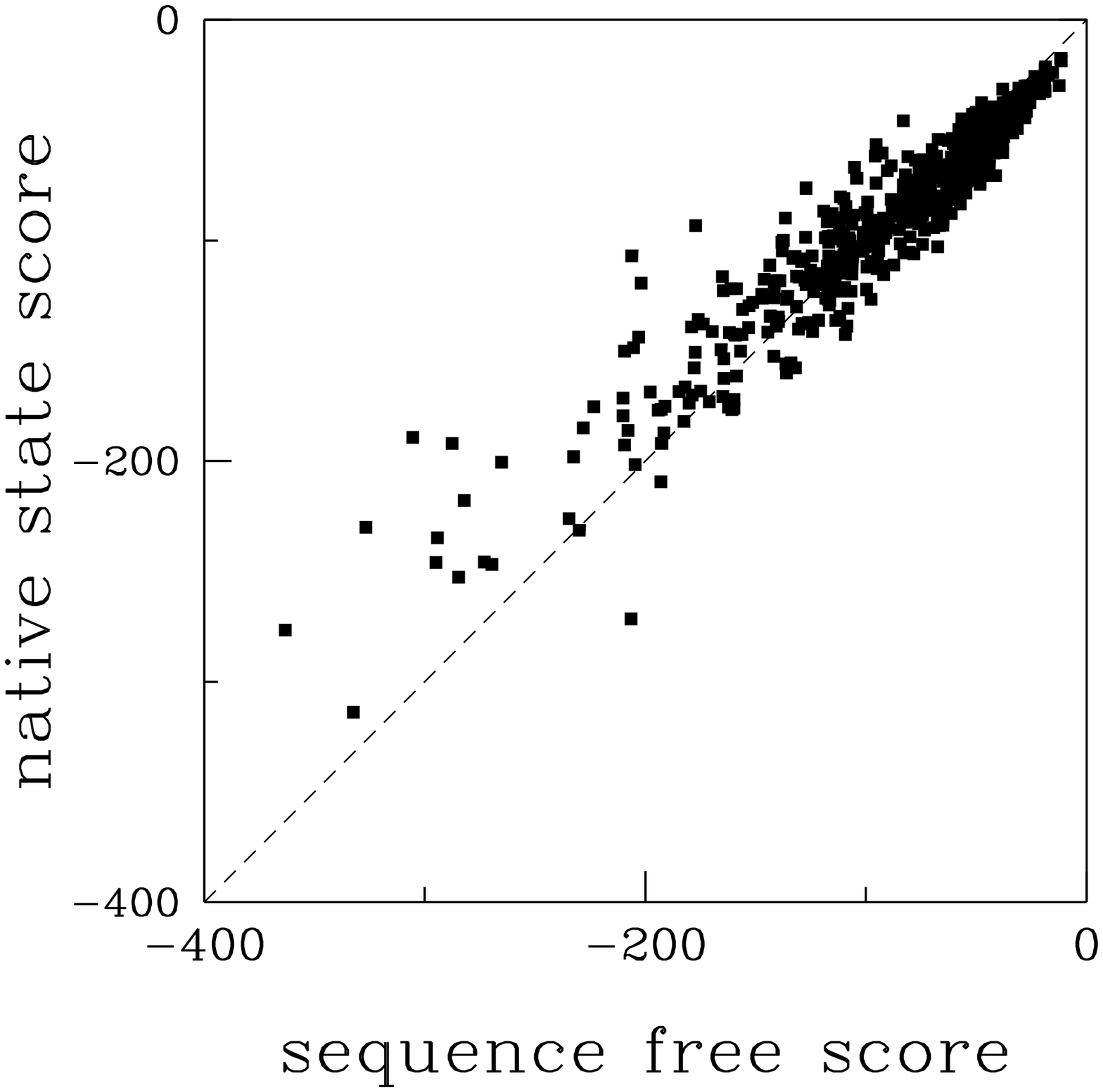}}
\vspace*{3.8cm}
\caption{ }
\end{figure}
%\vspace*{1cm}

%%FIGURE 4
%\begin{figure}
%\vspace*{3cm}
%\epsfxsize=5.2in
%\centerline{\epsffile{ncomodesa.eps}}
%\vspace*{2.8cm}
%\caption{ }
%\end{figure}
%%\vspace*{1cm}

%%FIGURE 5
%\begin{figure}
%\vspace*{3cm}
%\epsfxsize=5.5in
%\centerline{\epsffile{nnzvamo.eps}}
%\vspace*{2.8cm}
%\caption{ }
%\end{figure}
%%\vspace*{1cm}

\end{document}